\documentclass[12pt,preprint]{aastex}
\lefthead{Cunha et al.}
\righthead{$\omega$~Centauri}

\begin{document}

\title{Manganese Abundances in the Globular Cluster $\omega$~Centauri}

\author{Katia Cunha \& Verne V. Smith}
\affil{National Optical Astronomy Observatory, 950 N. Cherry Ave, Tucson, AZ, 85719 -  USA}

\author{Maria Bergemann}
\affil{Max-Planck Institute for Astrophysics, Karl-Schwarzschild Str. 1, 85741 Garching, Germany}

\author{Nicholas B. Suntzeff}
\affil{Texas A\&M University, Department of Physics and Astronomy \&
Mitchell Institute for Fundamental Physics and Astronomy, College Station, TX 77843-4242, USA}

\author{David L. Lambert}
\affil{University of Texas, 1 University Station, C1400, Austin, TX 78712 USA}

\begin{abstract}
We present manganese abundances in 10 red-giant members
of the globular cluster $\omega$ Centauri; 8 stars are from the most
metal-poor population (RGB MP and RGB MInt1) while two targets are members of the
more metal rich groups (RGB MInt2 and MInt3).
This is the first time Mn abundances have been studied in this peculiar stellar system.
The LTE values of [Mn/Fe] in $\omega$ Cen overlap those of Milky Way stars in
the metal poor $\omega$ Cen populations ([Fe/H]$\sim$-1.5 to -1.8), however unlike what is observed
in Milky Way halo and disk stars, [Mn/Fe] declines in the two more metal-rich RGB MInt2 and MInt3
targets. Non-LTE calculations were carried out in order to derive corrections
to the LTE Mn abundances.  The non-LTE results for $\omega$ Cen in comparison with
the non-LTE [Mn/Fe] versus [Fe/H] trend obtained for the Milky Way confirm and strengthen the
conclusion that the manganese behavior in $\omega$ Cen is distinct.
These results suggest that low-metallicity supernovae (with metallicities
$\le$ -2) of either Type II or Type Ia dominated the enrichment of the more metal-rich stars in $\omega$ Cen.
The dominance of low-metallicity stars in the chemical evolution of $\omega$ Cen has been noted
previously in the s-process elements where enrichment from metal-poor AGB stars is indicated.
In addition, copper, which also has metallicity dependent yields, exhibits lower values of
[Cu/Fe] in the RGB MInt2 and MInt3 $\omega$ Cen populations.
\end{abstract}


\section{Introduction}

The abundance ratios of chemical elements which have different nucleosynthetic origins 
can be used to constrain the history of chemical evolution in different types of stellar populations.
This is because stellar sources, such as massive stars that undergo core-collapse and explode
as supernovae of Types II or Ib/Ic, or binary systems that become Type Ia supernovae (SNe Ia), or asymptotic
giant branch (AGB) stars, produce and return to the interstellar medium different element abundance
ratios on differing timescales.   Abundance ratios of certain elements can thus be used
to explore which stars and in which proportions have contributed to chemical evolution within
stellar populations in galaxies. 

Of particular interest to chemical evolution of a stellar system are those elements
whose yields may depend on the metallicity of the progenitor star.  Metallicity-dependent
abundance patterns can retain `memory' of the metallicity distributions of their stellar
populations, regardless of how the overall metallicity of the parent galaxy evolves.   One important
element whose production may be metallicity dependent is manganese
(Z=25), which falls within the iron peak.  Manganese is produced in both core collapse supernovae (CC SNe) and 
SNe Ia, but the relative amounts from these two sources are not well constrained.
Manganese has one stable isotope, $^{55}$Mn;  this element is synthesized as a result
of explosive incomplete silicon burning (see, e.g., Thielemann et al. 2007 for a more complete discussion).

Woosley \& Weaver (1995), for example, model nucleosynthesis of massive stars exploding
as supernovae of Type II (SNe II) and find increasing Mn yields with increasing metallicity. Convolving the Mn
yields from the Woosley \& Weaver (1995) models with a Salpeter mass function leads to
values of [Mn/Fe] that decline steadily to -0.3 at [Fe/H]=-1 and then to -0.5 at [Fe/H]=-2. 
A decrease of [Mn/Fe] ratios with metallicity is reported in the LTE Mn I abundance analyses of 
Galactic stars such as Gratton (1989), Reddy et al. (2003; 2006), 
Johnson (2002), Cayrel et al. (2004), or Fetzing et al. (2007). Recent results from Bergemann \& Gehren (2008), however, 
indicate that the LTE approximation underestimates the abundances based on Mn I lines by 0.1 - 0.4 dex, 
with non-LTE effects being most pronounced in stellar atmospheres with low metal content. 
In the metallicity range investigated by Bergemann \& Gehren (2008), -2.5 $<$ [Fe/H] $<$ 0, 
the non-LTE [Mn/Fe] ratios in late-type stars are approximately solar. 
The flat [Mn/Fe] trend with [Fe/H] is in general agreement with the SNe II yields of Chieffi \& Limongi  (2004) 
who do not predict a strong depletion of Mn relative to Fe at low metallicities.

Manganese yields from SNe Ia models can span a range of values depending on the explosion mechanism, such as
slow or fast deflagrations, or deflagration-detonation events.  Iwamoto et al. (1999) investigate a variety of
models and find no large differences in Mn/Fe yields (i.e., less than a factor of $\sim$2) from models driven by
both slow and fast deflagrations, as well as deflagration-detonations, having different initial metallicities. 
Badenes et al. (2008) computed nucleosynthesis from 4 delayed detonation models, with different 
metallicities, as well as one deflagration model.  This group finds that Mn yields decline with decreasing
metallicity, despite the differences in the explosion mechanisms and initial conditions of the models.  

In addition to the results for the Galactic thin disk, thick disk and halo noted above, McWilliam et al. (2003) added
two other populations to the studies of manganese by measuring LTE [Mn/Fe] abundances for stellar
members of the Sagittarius dwarf spheroidal galaxy and stars from the Galactic bulge.  Both systems
exhibited somewhat different behaviors of [Mn/Fe] with [Fe/H], with the Sgr dwarf galaxy stars having
values of [Mn/Fe] that fall below the general LTE trend found for Milky Way disk or halo stars at a 
given [Fe/H].   Bulge stars exhibit opposite behavior, with their values of [Mn/Fe] falling above the
trend defined by the Milky Way disk and halo stars.  

The goal of this study is to add another distinct stellar population to analyses of manganese 
and its chemical evolution in different Galactic environments.  
$\omega$ Cen exhibits some peculiar characteristics in the nature of its chemical evolution, with 
perhaps the most striking being the large increase in the abundances of the heavy s-process elements 
(such as Ba or La) as the overall metallicity of cluster stars, measured by such elements as Fe, Ca, 
or Ti, increases (e.g. Norris \& Da Costa 1995).
$\omega$ Centauri, although historically classified as a globular 
cluster, is now thought possibly to be a surviving remnant of a captured small galaxy, with multiple
populations spanning a large range in metallicity (for a more detailed discussion see the review
by Smith 2004). Recently,  Carretta et al. (2010) have pointed out similarities between
the $\omega$ Cen populations and those from the Sagittarius dwarf galaxy. Five distinct stellar 
populations each with a different metallicity have been identified by Pancino et al. (2000) and 
Sollima et al. (2005). These studies label the distinct red giant branches from the $\omega$ Cen populations
as RGB MP (metal-poor); RGB MInt1; RGB MInt2; RGB MInt3 (intermediate metallicities); and RGB-a (anomolous,
with the highest metallicity).
 
Manganese abundances are presented here for the first time in $\omega$ Cen stars, with the sample 
consisting of 10 targets; 8 red giants are from the most metal poor populations (RGB MP and RGB MInt1)
and 2 stars are from the more metal rich RGB MInt2 and MInt3.  These $\omega$ Cen stars
have been analyzed in previous studies (Smith et al. 2000; Cunha et al. 2002), however Mn was
not included in the analysis.

\section{Observations and Stellar Parameters}

The high-resolution spectra analyzed in this study were obtained
using the cassegrain echelle spectrograph on the CTIO 4m Blanco
telescope. This data set is the same as analyzed in the
previous abundance study by Smith et al. (2000).  
The stellar parameters and microturbulent velocities adopted for the target stars
are also from this previous study (see Smith et al. 2000 for details). 
In brief summary, effective temperatures and microturbulence parameters were
found from the condition that the abundances derived from the LTE analysis of Fe I
lines of different excitation potentials and equivalent widths were equal.  In addition,
T$_{\rm eff}$s were estimated from the photometic indices (B-V) and (V-K).  Surface
gravities, and thus metallicities, were determined from the condition of ionization
equilibrium of Fe, i.e. that Fe I and Fe II lines computed under LTE yield equal abundances.
The adopted values of T$_{\rm eff}$, surface gravity (as log g), Fe abundances (A(Fe)),
and microturbulent velocities ($\xi$) for the $\omega$ Cen targets are listed in Table 1.
The  typical uncertainties are $\pm$100K in T$_{\rm eff}$, $\pm$0.3 dex in log g, $\pm$0.12
dex in A(Fe), and $\pm$0.3 km s$^{-1}$ in $\xi$. 
 
In addition to $\omega$ Cen stars, 4 M4 red giants were included
in this study. M4 is a well-studied Galactic globular cluster and provides
a comparison object for $\omega$ Cen. The M4 targets were observed with
the same echelle spectrograph on the CTIO 4m and these spectra were previously
analyzed by Drake et al. (1992), although Mn was not one of the elements
analyzed in that study. Table 1 includes the stellar parameters ($T_{\rm eff}$,
log $g$, $\xi$) and iron abundances for M4 stars taken from Drake et al. (1992).
 
\section{Manganese Abundance Analysis}

Manganese abundances were derived from LTE spectrum synthesis using the code MOOG 
(Sneden 1973). The model atmospheres adopted in this analysis are the same as
in Cunha et al. (2002) and are taken from the Bell et al. (1976) MARCS grid for metal-poor giants. 
The use of Bell et al. (1976) models is justified as comparisons to the current MARCS grid 
of model atmospheres (Gustafsson et al. 2008) indicate no measurable differences in the Mn 
I abundances for the range in effective temperatures and surface gravities of the studied stars.

The abundance of manganese in the target red giants were derived from Mn I lines near $\lambda$ 6000\AA. 
Three Mn I lines are available in this spectral region: $\lambda$$\lambda$ 6013.50\AA; 6016.64\AA; and 6021.80\AA.
In this study, only two Mn I transitions were analyzed as the Mn I line at 6016.64\AA\ 
is blended with a Fe I transition at 6016.604\AA. 
Our linelist in the 6015\AA\ region included this Fe I line with a gf-value from the Kurucz database 
(log gf= -1.82; from May et al. 1974 but renormalized to an average multiplet). 
However, the solar spectrum could not be adequately fit with the Kurucz gf-value
as the synthetic Mn I line at 6016\AA\ was clearly too broad when compared to the observed solar spectrum. 
Test calculations indicated that the Fe I line gf-value would have to be decreased significantly 
in order to properly match the width of the Mn I line in the solar spectrum. 
Given the uncertainties in the gf-value of the blending Fe I line, the Mn I line at 6016.64\AA\ 
was rejected from this study. 
 
The inclusion of hyperfine splitting (hfs) in the computation of synthetic spectra
is a requirement in order to  properly analyze the Mn I transitions, which are
affected by significant hyperfine splitting.  
The hfs data for both Mn I lines in this study were taken from Table 1 in Prochaska \& McWilliam (2000). 
We have recomputed the hyperfine structure of the $6013$\AA\ Mn I line using the magnetic dipole and 
electric quadrupole interaction constants for the upper level (e6S) of multiplet $16$ 
from Brodzinski (1987) and for the lower level (z6P*) from Handrich et al. (1969). 
The line equivalent widths of the $6013$\AA\ lines computed with our hfs and that of 
Prochaska \& McWilliam (2000) agree to within $1\%$.
The gf-values of Mn I $\lambda$ 6013.50\AA\ and 6021.80\AA\
are from Blackwell-Whitehead \& Bergemann (2007) and these are the same adopted
in Bergemann \& Gehren (2007).  
Table 2 provides the line list used in the abundance analysis of manganese. Additional atomic
lines from the Kurucz line list were added along with CN lines from Davis \& Phillips (1963;  which were
kindly  provided to us in digital form by A. McWilliam). 

Synthetic spectra were computed and manganese abundances were adjusted in order to best match the observed Mn I profiles.
The Uns\"old approximation (Uns\"old 1968) was adopted, without enhancements, for the van der Walls damping.
The Uns\"old interaction potentials are known to underestimate widths of strong metal lines 
in the solar spectrum (Gehren et al. 2001a,b), and another approach to compute the line broadening 
due to elastic collisions with H I atoms was developed by Anstee \& O'Mara (1995). 
However, for the parameters of stars in our analysis the differences in the profiles of Mn I lines 
of multiplet 16 computed with the Uns\"old or Anstee \& O'Mara (1995) formalisms are very small.
The spectral region containing the Mn I features is illustrated
in Figure 1 which shows synthetic and observed spectra for the target star
ROA383. As a comparison the Mn I lines were also synthesized 
in the well-studied giant Arcturus using the spectral atlas from Hinkle et al. (2000).
The stellar parameters adopted for Arcturus were $T_{\rm eff}$ = 4300 K, log $g$ = 1.7, 
and metallicity, [m/H] = $-$0.6 (Smith et al. 2000). This analysis of Arcturus
yielded a Mn abundance of A(Mn)= 4.70. The abundances of Arcturus are also included in
Table 1.

Uncertainties in the primary stellar parameters, T$_{\rm eff}$,
surface gravity, and the microturbulent 
velocity all affect the derived abundance.  
In order to estimate the errors in the Mn I abundances
introduced by uncertainties in stellar parameters, 
the following sensitivities to each variable were
calculated: ($\Delta$A$_{\rm Mn}$/$\Delta$T$_{\rm eff}$)$_{\rm -100K}$= -0.22,
($\Delta$A$_{\rm Mn}$/$\Delta$$\xi$)$_{\rm +0.3 km/s}$= -0.06, and 
($\Delta$A$_{\rm Mn}$/$\Delta$$log g$)$_{\rm +0.3}$= +0.02.  A quadrature
sum of these sensitivities results in estimated uncertainties in the manganese
abundances from the Mn I lines to be about $\pm$0.23, with temperature
errors being, by far, the dominant term.

\subsection {Non-LTE Effects}

The manganese abundance analysis in this study (and almost all previous studies of manganese in the 
literature) has adopted the simplifying assumption of LTE for the Mn I lines.  
Recently, however, non-LTE calculations of the Mn I lines have been done  
for the Sun (Bergemann \& Gehren 2007) and  for a sample of metal-poor
dwarfs and subgiants (Bergemann \& Gehren 2008).  Bergemann \& Gehren (2008) find that within the range 
of effective temperatures from 5000 - 6200K, surface gravities from 3.4 to 4.6 in log g, 
and metallicities from solar down to -3 in [Fe/H], the non-LTE corrections decrease with decreasing
T$_{\rm eff}$, increase with decreasing metallicity, and increase slightly
with decreasing gravity.  

Test non-LTE calculations were done for a few stars in our sample in order to 
investigate the influence of non-LTE effects on line formation of
the Mn I sample lines $\lambda$ 6013\AA\ and 6021\AA. Five targets were
selected in order to cover the range in stellar parameters and metallicities of the studied sample.
The statistical equilibrium calculations for Mn were performed with the
DETAIL code (Butler \& Giddings 1985) using a manganese model atom from Bergemann \& Gehren (2007). 
The model atom has three ionization stages containing 459 levels
and 2809 radiative transitions. Wavelengths and oscillator strengths are taken
from Kurucz \& Bell (1995). The cross-sections of inelastic collisions
with  H I atoms are computed with the Drawin's formula in the version of
Steenbock \& Holweger (1984) and multiplied by a scaling factor of 0.05. This value
was found from the analysis of Mn I lines in the solar and stellar
spectra (Bergemann \& Gehren 2007; 2008). Photoionization
cross-sections were computed from the Kramer's formula using effective principle
quantum numbers. The Mn model is complete up to the Mn III ground state.
This is important for cool low-gravity atmospheres, because the number densities
of neutral and singly-ionized atoms at line formation depths are comparable and
statistical equilibrium of Mn is established by non-LTE processes in the atoms of
both ionization stages.
The model atmospheres for Mn statistical equilibrium calculations in this work 
were computed by Grupp (2010; private communication)
with the code MAFAGS-OS  (Grupp 2004 a, b). These models are standard plane parallel static LTE models,
and they are virtually identical to the respective model atmospheres used in the LTE analysis;
in particular, the temperature stratifications agree to within $50$K  at $-2.5 <  log$ $\tau < 1$ (see Figure  2).
Since non-LTE effects also depend on the elemental abundance, the statistical equilibrium calculations were
performed with Mn abundances derived from the LTE analysis.

The detailed discussion of non-LTE equilibria in Mn for solar-type stars was given
in Bergemann \& Gehren (2007; 2008). The character of interaction processes 
does not change for stellar parameters, which form the basis of the current analysis. 
The diagrams of level departure coefficients
\footnote{Departure coefficients characterize deviations of number densities of
atoms in the level $i$ from their LTE values, $b_i = n_i^{\rm non-LTE}/n_i^{\rm
LTE}$} for two model atmospheres from our sample are shown in Figure 3.
Since in this study we are solely interested in investigating the formation of the
Mn I  lines of multiplet 16, 
only selected levels are indicated. The Mn I levels
are underpopulated at optical depths $log$ $\tau (500 nm) < 0.7$, $b_{i} < 1$, which is due to
line pumping (see Bergemann \& Gehren 2007) combined with overionization
from intermediate-excitation Mn I levels. Overionization from
densely-populated low-excitation Mn I levels, which is the dominant non-LTE
effect in solar-type stars at sub-solar metallicities, is not very efficient at the
low temperatures encountered in the atmospheres of giants. Also, collisional
interaction of the majority of levels is very weak due to low densities and
metallicities of the models. This is manifested in a rather irregular behavior
of departure coefficients of Mn I  levels with depth, as seen in Figure 3.

The departure coefficients were used to compute non-LTE abundance corrections
$\Delta_{\rm non-LTE}$ (defined as the difference in abundances required to fit 
non-LTE and LTE profiles, $\Delta_{\rm non-LTE}$= A$_{\rm non-LTE}$ - A$_{\rm LTE}$)  
to the lines of multiplet $16$ for five stars from our sample.  
The non-LTE corrections obtained for these stars are found in Table 3. 
$\Delta_{\rm non-LTE}$ are positive and range from $+0.05$ dex for the models with higher
metallicities to $+0.3$ dex for the model atmosphere of ROA 213
with [Fe/H] $= -2$. 
The variations of  $\Delta_{\rm non-LTE}$ between models can be understood 
from the inspection of departure coefficients, which define the behavior of line 
opacities $\kappa_\nu^{l} \sim b_i$ and
source functions $S^l \approx \frac{b_j}{b_i} B_{\nu}^l$, where $B_{\nu}^l$ is
the Planck function and indices $j$ and $i$ refer to the upper and lower level of a
transition, respectively. At [Fe/H] $\sim$ -1, both Mn I lines of
multiplet 16 are strong enough and their non-LTE corrections are determined by
the competition of $S^{l} > B_{\nu}^l$ in the wings and $S^{l} < B_{\nu}^l$ in the
core. The latter condition implies a strengthening of a line, thus non-LTE
corrections are small. At a decreased metallicity, [Fe/H] $= -2$, the lines
become weak and are formed in the deeper layers, where $S^{l} > B_{\nu}^l$, and
also $b_{i} < 1$. The raised non-LTE source function and decreased opacity cause
significant weakening of the line, and, thus, large positive non-LTE abundance
corrections.  
 
Thus, for the range of stellar parameters investigated here, the magnitude of the non-LTE
abundance corrections for both Mn I lines of multiplet 16 is determined primarily by metallicity.
$\Delta_{\rm non-LTE}$ are largest ($\sim$ 0.3 dex) for the most metal-poor $\omega$ Cen giant ([Fe/H] =
-2), while for the red giants with larger metallicity ([Fe/H] = -1) and low Mn abundances,
$\Delta_{\rm non-LTE}$ $\sim$ 0.1 dex

\section{Discussion}

The final LTE manganese abundances that are shown in Table 1 are combined with iron abundances (from
Smith et al. 2000 for $\omega$ Cen and Drake et al. 1992 for M4) and compared with abundances
from other stellar samples.  
The discussion begins with a comparison of Mn abundances in $\omega$ Cen
with stars from the thin disk, thick disk, and halo, as well as Sgr dwarf galaxy members.  This is then
followed by a section that compares $\omega$ Cen with those abundances from a large sample of Milky
Way globular cluster giants and field stars.  The discussion concludes by highlighting how the
manganese abundances provide additional insight into chemical evolution within $\omega$ Cen.

\subsection{[Mn/Fe] in $\omega$ Cen, Milky Way Field Stars and the Sagittarius Dwarf Galaxy}

The initial comparison of $\omega$ Cen stars with other stellar populations is shown in Figure 4, where
the LTE abundance ratios of Ca/Fe and Mn/Fe, computed as A(Ca or Mn) - A(Fe), are shown versus 
A(Fe), with
Ca/Fe plotted in the top panel and Mn/Fe in the bottom panel. It is of interest to discuss calcium results
first, as Ca is a well-studied $\alpha$-element whose yields are not expected to be metallicity dependant.
The solar values are indicated by the solar
symbols and abundances plotted this way can easily be transformed to values of [Ca/Fe] and [Mn/Fe] by using
the absolute solar values (A(Ca)$_{\odot}$=6.34, A(Mn)$_{\odot}$=5.43, and A(Fe)$_{\odot}$=7.45;
Asplund et al. 2009).  
The Milky Way field-star samples are plotted as blue small symbols and are 
taken from Reddy et al. (2003; 2006; open circles); 
Fulbright (2002; open squares); Johnson (2002; open squares); Cayrel et al. (2004;
open triangles) and Mcwilliam et al. (1995; open pentagons).  
We also added the results for metal-rich $\omega$ Cen stars by Pancino et al. (2002; red filled squares)
and the Sagittarius dwarf results by Mcwilliam et al. (2003) and Sbordone et al. (2007) are
represented as green asterisks.

Focussing first on LTE Ca/Fe in the field stars, the overall 
trends are clear, with values of Ca/Fe increasing with decreasing Fe abundance, 
which befits calcium's classification as an $\alpha$-element (Figure 4; top panel).
Note that in this figure, no distinction is made between thin and thick disk populations,
which may account for some of the scatter in Ca/Fe at disk metallicities.
In this panel, the $\omega$ Cen members follow roughly the LTE trend defined by the 
Milky Way field stars (as do the Sgr dwarf galaxy stars, although the most metal-rich 
Sgr stars tend to fall below the Ca/Fe values defined by the Galactic field stars).  
Note that the histogram plotted within the top panel of Figure 4 represents the 
Fe-abundance distribution function derived for $\omega$ Cen by Sollima et al. (2005) and 
is included as a  schematic representation of where the $\omega$ Cen stars studied here, along with those
from Pancino et al. (2002) fall within the overall metallicity distribution of the cluster.

Turning now to the bottom panel, with Mn/Fe plotted versus A(Fe) (both quantities being LTE
values), the trend for the Milky Way field stars
can be described as one in which Mn/Fe declines steadily with decreasing Fe abundance, to values of 
$\sim$ -0.4 dex below solar at iron abundances about 1/10 solar.  The ratio then remains fairly constant, 
albeit with considerable scatter, which may be due to systematic effects between different studies,
as metallicity continues to decrease.  There may be another downturn
in Mn/Fe at the very lowest metallicities, however the non-LTE corrections found by Bergemann \& Gehren (2008)
predict increasingly large corrections as the metallicity declines, so part of this apparent decline
at very low Fe abundance may be due to non-LTE effects, such that the value of Mn/Fe is more nearly constant.  
As noted by McWilliam et al. (2003), the Sgr stars included in Figure 4 (bottom panel) fall below the Mn/Fe 
versus Fe trend set by the Milky Way stars, with the offset being $\sim$0.2 dex.  

With the inclusion of $\omega$ Cen Mn/Fe ratios,
the differences between the Milky Way populations become even more extreme than Sgr, with a decline in
Mn/Fe as Fe increases in $\omega$ Cen.  In the most metal-rich $\omega$ Cen stars in this sample
the values of Mn/Fe fall about 0.6-0.7 dex below the LTE trend established for the field stars.  This offset
between $\omega$ Cen and Milky Way stars is reminiscent of what was found for Cu/Fe by Cunha et al.
(2002) for $\omega$ Cen, and was also found for Cu/Fe in Sgr by McWilliam \& Smecker-Hane (2005).
The chemical evolution of manganese in $\omega$ Cen is yet another observation that implicates $\omega$ Cen
as a chemically distinct population when compared to the Milky Way and that can be more 
closely associated with abundance patterns in small galaxies at comparable
metallicities. We note, however, that in $\omega$ Cen there are significantly more metal-rich stars 
(e.g. Norris \& Da Costa 1995; Pancino et al. 2002) that reach near-solar metallicities. 
So far these stars have not yet been analyzed for Mn. 

As discussed in Section 3.1, the non-LTE Mn results indicate it is unlikely that the low ratios of Mn/Fe
observed in $\omega$ Cen result from non-LTE effects.  In fact, a direct application of the non-LTE corrections
to the $\omega$ Cen stars would increase the differences in Mn/Fe between the metal-poor and metal-rich
members of $\omega$ Cen, as shown in the top panel of Figure 5.  

One can also compare non-LTE Mn abundances in Milky Way field stars with the non-LTE results for the $\omega$ Cen
stars derived here.  Bergemann \& Gehren (2008)  provide the Milky Way field-star sample by having analyzed 
14 stars spanning a range in [Fe/H] from $\sim$ 0.0 to -2.5.  In this case, a comparison of $\omega$
Cen with the field stars leads to an even more striking difference, with Bergemann \& Gehren (2008) finding
essentially constant values of [Mn/Fe] not very different from solar across the observed range in [Fe/H]: the mean and
standard deviation are [Mn/Fe]=  -0.05$\pm$0.11, with no significant trend found with metallicity.  The $\omega$
Cen stars fall well below these values, with Mn/Fe decreasing slightly with increasing metallicity.
Differences between $\omega$ Cen stars
and the Milky Way field stars exist whether the comparison is with LTE or non-LTE abundances.

There still remains concern that perhaps ratioing non-LTE Mn I abundances with LTE Fe I
abundances might induce spurious values of Mn/Fe so as to perturb significantly the
location of the $\omega$ Cen giants in the Mn/Fe -- A(Fe) plane.  Since non-LTE 
calculations of Fe I and Fe II in cool red giants remain questionable, the test
of Mn I to Fe can be expanded to include abundances from Fe II (since in these red
giants virtually all iron is in the form of Fe I + Fe II).  Such a test is possible
to carry out by using results from Smith et al. (2000), who included both Fe I and
Fe II in their analysis.

Smith et al. (2000) forced equal abundances from both Fe I and Fe II as a way of
setting the surface gravities, although another method
to derive surface gravities is to use the rather well-defined luminosities of the
$\omega$ Cen red giants (which are in Smith et al. 2000 Table 1) in combination
with T$_{eff}$ and stellar mass, M.  The luminosities
in Smith et al. were derived by using a distance modulus (m-M)=13.6 and an E(B-V)=0.11 with 
K-magnitudes to determine M$_{\rm K}$.  Bolometric corrections were then taken from
Bessel et al. (1998) to arrive at M$_{\rm bol}$.  
Since L $\alpha$ R$^{2}$T$_{\rm eff}$$^{4}$ and g $\alpha$ MxR$^{-2}$, then
g $\alpha$ MxL$^{-1}$xT$_{\rm eff}$$^{4}$.  With known T$_{\rm eff}$ and L, these can
be combined with the rather restricted range of masses allowed for old, low-mass red
giants to derive ''evolutionary'' gravities; in this case, we use 0.8M$_{\odot}$ for
the masses of the $\omega$ Cen red giants.  The evolutionary gravities can then be
compared to the spectroscopic gravities to search for systematics or large scatter
that would lead to large differences in the LTE Fe I and Fe II abundances.  Such a
comparison for the 10 $\omega$ Cen red giants finds no large differences, with the
result that the mean and standard deviation of
$\Delta$$_{\rm log g}$(evolutionary log g - spectroscopic log g)= +0.11$\pm$0.18 dex.
This small offset and scatter is within the uncertainties in the spectroscopic gravities
themselves, as well as the masses.  Employing the evolutionary gravities would lead
to Fe II abundances that are only slightly larger, in the mean, by +0.12 dex.  Using
various combinations of spectroscopic or evolutionary gravities and either Fe I or 
Fe II abundances, will not alter the conclusion that the metal-rich $\omega$ Cen 
stars exhibit low values of Mn/Fe when compared to other stellar populations.
The low values of Mn/Fe in the more metal-rich $\omega$ Cen stars, relative to Milky Way field stars, is a real effect
that needs to be explained by chemical evolution within this peculiar stellar system.

\subsection{[Mn/Fe] in $\omega$ Cen and Milky Way Globular Clusters}

It has been previously shown that the general pattern of globular cluster
abundances (with very few exceptions) typically follows the trends observed for the disk and halo at
a given iron abundance. 
Sobeck et al. (2006) derived Mn abundances in more than 200 giant stars in 19 Milky Way globular 
clusters, along with about 200 field giants.  Their results for [Mn/Fe] versus [Fe/H] revealed no significant
differences between the globular clusters and field populations.

A comparison of our derived LTE abundances for $\omega$ Cen and M4 with the Sobeck et al. (2006) 
LTE results 
are of particular interest as globular cluster giants in their sample have similar 
stellar parameters (T$_{\rm eff}$ and log g) to the target stars studied here. 
It should be noted that
both studies analyzed the same Mn I lines although Sobeck et al. adopted the gf-values from 
Booth et al. (1984), which are higher than the ones adopted here.
Figure 5 (bottom panel) illustrates a comparison between the Mn/Fe versus A(Fe) abundance results in 
the two
studies; the Sobeck et al. (2006) abundances in the figure were adjusted in order to account for differences
in the adopted gf-values.  Since the stellar parameters in Sobeck et al. (2006) are similar to those
for the $\omega$ Cen and M4 stars studied here, it is likely that non-LTE corrections to their results
will be similar to those computed here.  The trend of increasing positive non-LTE abundance corrections
for the Mn I lines with decreasing metallicity would likely apply to the Sobeck et al. (2006) results.
Whether comparing LTE abundances or likely non-LTE abundances, as with Figure 4 (bottom panel), 
this demonstrates that $\omega$ Cen is different 
from the typical Milky Way globular cluster population in its low values of Mn/Fe.  

To reinforce the differences found in values of Mn/Fe in the more
metal-rich $\omega$ Cen red giants, when compared to other globular
cluster red giants, the results derived here for
the 4 M4 targets are also plotted in the bottom panel of Figure 5 (as filled
green squares).  Recall that these M4 red giants were observed with
the same spectrograph and their analysis was the same as that used for
the $\omega$ Cen sample.  The location of the M4 abundances derived 
for Mn/Fe and A(Fe) falls within the scatter of results from the 
Sobeck et al. (2006) globular cluster abundances.
In fact, two of the M4 stars analyzed here are in the
Sobeck et al. (2006) study and the black lines from these two M4 stars
connect the abundances derived in this study to those values from Sobeck et al. (2006):
the agreement between the two studies is very good.  The two most
metal-rich $\omega$ Cen red giants remain depressed in their Mn/Fe
abundance ratios when compared to stars of similar metallicity in M4. 

\subsection{Manganese and the Stellar Populations of $\omega$ Cen}

Having established low Mn/Fe ratios in the two
most metal-rich $\omega$ Cen stars analyzed here, it is of interest to place
these depressed Mn/Fe ratios in the context of the star formation history
of $\omega$ Cen and its chemical evolution.
As mentioned in the introduction, large-field photometric surveys, 
such as those of Lee et al. (1999),
Pacino et al. (2000), or Sollima et al. (2005), which include tens of
thousands, or up to 100,000 stars, have identified discrete red giant
branches (RGBs) in color-magnitude diagrams (CMDs) in, for example,
B versus B-V or I versus B-I.  Such discrete RGBs represent stellar
populations, each having a reasonably well-defined metallicity and 
probably age.

Sollima et al.(2005) identified 5 distinct RGBs, which can be ordered by
metallicity.  These RGBs have been labelled as MP (for metal poor),
then 3 intermediate-metallicity branches ordered in increasing
metallicity as MInt1, MInt2, and Mint3, and finally a quite metal-rich
group called RGB-a ('a' for anomalous).  These separate RGBs are
clearly visible in the Sollima et al. (2005) MDF plotted schematically
in the top panel of Figure 4. 

Sollima et al. (2005) define metallicity limits for each of the RGBs
and their definition can be used to associate each of the $\omega$ Cen
red giants here with one of the branches.  The last column of Table 1
lists this population classification.  Of course there is uncertainty
in this classification due to both uncertainties in the Fe abundances
from Smith et al. (2000) and to overlap in neighboring RGBs.  Uncertainties
in A(Fe) are $\sim$$\pm$0.1 dex and the RGB separations are about 0.2
dex, so the classification uncertainties are expected to be confined to
a neighboring RGB.  Within this classification scheme, the two stars
having significantly low Mn/Fe ratios (ROA219 and ROA324) are members
of MInt2 and MInt3.  The MInt2 and MInt3 populations are part of the
extended metal-rich tail, with most (if not all) of the Fe arising from
CC SNe enrichment.  In addition, these stars exhibit large s-process
enhancements, which indicates substantial enrichment from low-metallicity
AGB stars (Smith et al. 2000).  Pancino et al. (2002) analyzed iron, copper and
$\alpha$-elements and found no
measurable SN Ia enrichment in 3 stars from the MInt3 and MInt2
populations, but did find such enrichment in 3 RGB-a stars (their
Ca/Fe ratios are plotted in the top panel of Figure 4).

\subsubsection{The effect of He enhancements on the Mn Abundances}

In addition to the peculiar heavy-element chemical evolution in $\omega$ Cen,
the presence of a blue main-sequence (Bedin et al. 2004) that is more
metal-rich than the red main-sequence (Piotto et al. 2005) indicates significant
He-enrichment may accompany the overall metallicity increase.  It has been
estimated by Norris (2004) that a He overabundance of $\Delta$Y $\sim$ +0.15
is required to explain the blue main-sequence stars. 
If this helium enhancement is correct, it suggests that some stars in this sample
may have undergone some degree of He enrichment, so it is worth investigating
if this might affect significantly Mn/Fe ratios.  

Bohm-Vitense (1979) discussed
the effects that enhanced He abundances would have on the strengths of
spectral lines in late-type stellar atmospheres and her discussion can be
used to estimate how an increased value of Y might affect, for example Mn/Fe
ratios.  Using her 'case b' (where electrons for H$^{-}$ arise from metals
which are partially ionized or mostly neutral, as expected in low temperature
red giants) for neutral atomic lines in an atmosphere where the dominant
continuous opacity is H$^{-}$, the dependence on helium abundance is small
(line strength $\alpha$ (1 + 4y)$^{1/6}$(1 + y)$^{2/3}$, where y = N(He)/N(H)).
If the extremes in the $\omega$ Cen populations are Y$\sim$0.25 and 0.40
(Sollima et al. 2008, in particular their Figure 2), the estimated effects
on neutral metal lines from Bohm-Vitense's (1979) analysis would be small,
y changing from 0.08 to 0.17 in her derived equation.  In addition, the
peculiarity of the $\omega$ Cen stars is in the ratio of Mn/Fe, which is
dependent on Mn I and Fe I lines, so effects due to enhanced He abundances
would cancel to some degree.  Bohm-Vitense (1979) also point out that, for
this case of a cool atmosphere, He effects would be largest for lower
ionization species and she notes that Ca I lines would be affected more
than, say Fe I.  Reinspection of the top panel of Figure 4 reveals that
Ca/Fe in $\omega$ Cen follows the trends defined by other stellar populations,
suggesting that any He enhancements are probably not affecting significantly
the derived metal abundances. 

Finally, we have tested the effect of He-enhancement on the derived abundances
by computing the Mn I line formation for the model atmosphere of the giant M4 2617 with $\Delta$ ${\rm Y} = +0.1$. 
The temperature stratification of this model is shown in Figure 2. 
We find that the profiles of the 6013\AA\ and 6021\AA\ Mn I lines computed 
with $\Delta$ $\rm{Y} = +0.1$ model are almost identical to those computed with the model 
atmosphere with normal He abundance.

\subsection{Chemical Evolution within $\omega$ Cen as Revealed by Low Manganese Abundances}

Recently, Cescutti et al. (2008) modelled the chemical evolution of manganese in three distinct stellar
populations: Milky Way field stars (the Solar neighborhood, containing thin and thick disk members, as
well as halo stars), the
Bulge, and Sgr dwarf galaxy members.  These three populations exhibit different behaviors, which can be
summarized briefly as, at a given Fe abundance, Mn/Fe ratios in the Bulge fall above the Solar neighborhood
stars, while Mn/Fe ratios in Sgr fall below the Solar neighborhood values.  Note that all of these abundance
studies used by Cescutti et al. (2008) are based on LTE analyses.  Their results must be interpreted with
caution, as non-LTE calculations need to be carried out in order to confirm or not the model conclusions. 

Cescutti et al. (2008) use a model for the Solar neighborhood from
Fran\c cois et al. (2004) and Chiappini et al. (1997), which assumes two Galactic accretion events, the first
of which formed the halo and thick disk, while the second formed the thin disk.  The infalling gas in the
accretion events is taken to be primordial.  For the Bulge, Cescutti et al. use a model with a rapid
formation time (0.3-0.5 Gyr) and very efficient star formation (with the star formation per unit mass
of gas being some 20 times larger than for the Solar vicinity).  This rapid formation and vigorous star
formation in the Bulge explains the large values of [$\alpha$/Fe] found in bulge stars by a number of
studies (Cunha \& Smith 2006; Zoccali et al. 2006; Lecureur et al. 2007; Fulbright et al. 2007; 
Mel\'endez et al. 2008, Ryde et al. 2010).  
In Sgr, it is assumed that  stars formed over several Gyrs, with
strong galactic winds driven by supernovae leading to significant gas loss.  This galactic wind results
in an overall decrease in star formation efficiency and explains the low values of [$\alpha$/Fe] observed
in the more metal-rich Sgr members.
When the above chemical evolution models are applied to tracking the chemical evolution of manganese,
Cescutti et al. (2008) find that the best fits to the three trends of Mn/Fe versus A(Fe) defined by
each distinct population result when Mn is produced via metallicity-dependent yields in SN Ia.

If the Cescutti et al. (2008) model is applied to $\omega$ Cen, the implication is that metal-enrichment
at the metal-rich end of the distribution in this stellar system was driven by metal-poor SN Ia.  
This is a possibly viable
model, as the majority of stars in $\omega$ Cen are from the most metal-poor group (as evidenced by the
histogram in the 
top panel of Figure 4), thus the time-delay of SN Ia to the chemical evolution would result in the metal-poor
SN Ia systems contributing to the chemical makeup of the more metal-rich stars.  Such SN Ia systems would
be efficient producers of Fe, but would not synthesize much Mn, leading to low values of Mn/Fe.  Quantitative
predictions of Mn/Fe ratios await a detailed star formation history, probably with significant galactic
winds, coupled to a chemical evolution model.

Such a model for $\omega$ Cen has already been developed by Romano et al. (2007) and then applied by
Romano \& Matteucci (2007) to explain the low values of Cu/Fe found by Cunha et al. (2002) and
Pancino et al. (2002).  This model assumes that $\omega$ Cen began its history as a dwarf spheroidal galaxy
that evolved initially in isolation and whose chemical evolution was affected strongly by galactic
winds (as with Sgr).  The conclusion by Romano \& Matteucci (2007) concerning copper, was that it was
produced via a metallicity-dependent process in massive stars; in this case it is the weak s-process
occurring during He-burning in these stars, with the neutron source being $^{22}$Ne($\alpha$,n)$^{25}$Mg.
Since the amount of $^{22}$Ne depends on the abundance of $^{14}$N, whose value depends on stellar
metallicity, this renders the strength of the s-process proportional to metallicity.

Our result for $\omega$ Cen is that Mn behaves as a metallicity-dependent element, in a similar manner to that
of Cu, and may provide further constraints on both chemical evolution within this peculiar system, as well as
the origins of both Mn and Cu. Note that Wylie de Boer et al. (2010) used low values of Cu/Fe as one
indicator that stars in Kapteyn's group may represent an $\omega$ Cen stream. Our results here
suggest that Mn/Fe would also be a useful tracer of possible $\omega$ Cen tidal streams.
 
\section{Conclusions}

Manganese abundances have been measured for the first time in the peculiar
globular cluster $\omega$ Cen, with the analysis of 10 red giants spanning
a range in metallicity from [Fe/H]= -1.9 to -0.9.  The analysis is based
on Mn I lines using both LTE and non-LTE calculations to derive abundances. 
In addition, the possible effects of enhanced He abundances on derived Mn 
abundances were investigated for the more metal-rich $\omega$ Cen giants
and were found to be negligible.

The novel result is that two members from the more metal-rich populations
with $\omega$ Cen (RGB MInt2 and MInt3) exhibit low ratios of Mn/Fe in
comparison to Galactic field stars, as well as other globular cluster stars
at the same metallcitiy ([Fe/H]$\sim$-1).  Differences between $\omega$ Cen
and the other Milky Way populations exist whether the comparison is made
using LTE or non-LTE abundances.  The low abundances of Mn may indicate that
low-metallicity progenitors to supernovae (of either core collapse 
or SNe Ia) dominated the production of manganese within $\omega$ Cen.  This
result for Mn is similar to what has been noted previously for the
behavior of copper (which in some nucleosynthesis processes has 
metallicity-dependent yields) in $\omega$ Cen (Cunha et al. 2002).

The behavior of Mn in the more extreme metal-rich $\omega$ Cen population
(RGB-a) remains to be probed.  In future studies it would be of interest
to determine Mn abundances in the most metal-rich $\omega$ Cen population;
for example, by analyzing Mn in the 3 most metal-rich red giants studied 
to date (ROA 300, WFI22068, and WFI222679) by Pancino et al. (2002).  The
behavior of the manganese abundances in these more metal-rich stars will
provide further insight into both the origins of Mn and the nature of
star formation within $\omega$ Cen during the final throes of its chemical
evolution.

MB thanks Dr. Frank Grupp for providing MAFAGS-OS model atmospheres for
selected stars. This research was supported in part by the National Science Foundation 
(AST 06-46790 to KC and VVS).  DLL thanks the Robert A. Welch Foundation
for support via grant F-634.

%\clearpage

\clearpage

\begin{deluxetable}{ccccccc}
\setcounter{table}{0}
\tablewidth{400pt}
\tablecaption{Program Stars and LTE Abundances}
\tablehead{
\colhead{Star} &
\colhead{T$_{\rm eff}$(K)} &
\colhead{Log g} &
\colhead{$\xi$ (km-s$^{-1}$)} &
\colhead{A(Fe)} &
\colhead{A(Mn)} &
\colhead{RGB Type}
}
\startdata
ROA102 &  4400 & 1.0  & 3.0 & 5.69 & 3.23 & MP \\
ROA209 &  4500 & 1.2  & 2.0 & 5.72 & 3.36 & MP \\
ROA213 &  4500 & 1.0  & 1.9 & 5.53 & 2.98 & MP \\
ROA219 &  3900 & 0.7  & 1.7 & 6.25 & 3.34 & Mint2 \\
ROA236 &  4200 & 0.7  & 1.7 & 6.06 & 3.79 & MInt1 \\
ROA238 &  4550 & 1.2  & 1.7 & 5.72 & 3.14 & MP \\
ROA245 &  4300 & 0.7  & 1.8 & 6.10 & 3.42 & MInt1 \\
ROA253 &  4300 & 0.7  & 1.9 & 6.08 & 3.50 & MInt1 \\
ROA324 &  4000 & 0.7  & 1.9 & 6.57 & 3.64 & MInt3 \\
ROA383 &  4400 & 1.0  & 1.8 & 5.89 & 3.35 & MInt1 \\

M4 2519 & 4480 & 1.3  & 1.8 & 6.27 & 3.85 & ... \\
M4 2617 & 4280 & 1.2  & 1.5 & 6.32 & 3.95 & ... \\
M4 3624 & 4300 & 1.3  & 1.5 & 6.28 & 3.99 & ... \\
M4 3612 & 4350 & 1.3  & 1.5 & 6.39 & 4.08 & ... \\
$\alpha$ Boo & 4300 & 1.7 & 1.6 & 6.78  &  4.70 & ... \\
\enddata

\end{deluxetable}

\clearpage

\begin{deluxetable}{cccc}
\setcounter{table}{1}
\tablewidth{400pt}
\tablecaption{Linelist for the Mn I line regions}
\tablehead{
\colhead{$\lambda$ (\AA)} &
\colhead{Species} &
\colhead{$\chi$ (eV)} &
\colhead{Log gf}
}
\startdata
6013.020 & Mo I                                &   3.40  & -0.955 \\
6013.166 &  C I                                &   8.65  & -1.370 \\
6013.168 &  $^{12}$C$^{14}$N   &    0.77 &  -2.798 \\
6013.187 &  $^{12}$C$^{14}$N   &   1.87  & -1.760 \\ 
6013.206 &  Cr II                              &   8.42  & -3.522 \\
6013.213 &  C I                                &   8.65  & -1.470 \\
6013.218 &  $^{12}$C$^{14}$N   &   0.81  & -2.305 \\
6013.272 &  $^{12}$C$^{14}$N   &   1.76  &  -1.533 \\
6013.293 &  V I                                &   4.13  & -4.513 \\
6013.347 & Fe II                              &   8.07  &  -3.485 \\
6013.383 & Cr I                               &   5.61  &  -3.938 \\
6013.417 & Ti I                              &  1.07   & -3.100 \\
6013.420 & Ce I                              &   0.30  &  -0.526 \\
6013.478 & Mn I                              &   3.07  & -0.946 \\
6013.499 & Mn I                              &   3.07  & -1.158 \\
6013.518 & Mn I                              &   3.07  & -1.431 \\
6013.527 & Mn I                              &   3.07  & -1.635 \\
6013.533 & Mn I                              &   3.07  &  -1.841 \\
6013.538 & Mn I                              &   3.07  & -1.489 \\
6013.547 & Mn I                              &   3.07  &  -1.510 \\
6013.550 & $^{12}$C$^{14}$N    &   0.82  &  -2.014 \\
6013.552 & Mn I                              &   3.07  &  -1.665 \\
6013.556 & $^{12}$C$^{14}$N    &   0.76  &  -2.596 \\
6013.562 & Mn I                              &   3.07  &  -1.987 \\
6013.566 & Mn I                              &   3.07  &  -2.033 \\
6013.567 & Mn I                              &   3.07  &  -2.589 \\
6013.568 & Mn I                              &   3.07  &  -2.209 \\
6013.640 & Co I                              &   4.48  &  -2.870 \\
6013.734 & V II                              &   6.52  &  -4.633 \\
6013.773 & $^{12}$C$^{14}$N   &   1.70  &  -1.922 \\
6013.962 & Ti II                             &   8.11  &  -2.575 \\
         &                    &         &         \\
6021.118 & $^{12}$C$^{14}$N   &  1.89  &  -1.750 \\
6021.123 & Gd I                             &   0.84 &  -0.992 \\
6021.129 & $^{13}$C$^{14}$N   &   0.81  &  -2.098 \\
6021.165 & Ni I                               &   5.28 &  -3.539 \\
6021.275 & Cr I                               &   3.85 &  -2.593 \\
6021.276 & Ca I                              &   5.58 &  -4.150 \\
6021.299 & $^{13}$C$^{14}$N   &   1.65  &  -1.600 \\
6021.525 & W I                               &   2.24  &  -1.430 \\
6021.526 & $^{12}$C$^{14}$N   &  0.76  &  -2.982 \\
6021.660 & Sc I                              &   2.11 &  -2.020 \\
6021.710 & $^{12}$C$^{14}$N   &  0.78  &  -2.216 \\
6021.739 & Sm II                            &  1.38  &  -1.745 \\
6021.746 & Mn I                             &  3.08  &  -2.822 \\
6021.764 & Ca I                             &  5.59  &  -4.783 \\
6021.772 & Mn I                             &  3.08  &  -1.605 \\
6021.774 & Mn I                             &  3.08  &  -2.470 \\
6021.787 & Fe I                              & 2.20   &  -4.253 \\
6021.795 & Mn I                             & 3.08   &  -1.429 \\
6021.798 & Mn I                             & 3.08   &  -2.345 \\
6021.804 & Mn I                             & 3.08   &  -0.687 \\
6021.813 & Mn I                             & 3.08   &  -1.403 \\
6021.817 & Mn I                             & 3.08   &  -2.425 \\
6021.820 & Fe I                              & 4.28   &  -4.109 \\
6021.821 & Mn I                             & 3.08   &  -0.827 \\
6021.827 & Mn I                             & 3.08   &  -1.470 \\
6021.834 & Mn I                             & 3.08   &  -0.985 \\
6021.837 & Mn I                             & 3.08   &  -1.646 \\
6021.838 & $^{13}$C$^{14}$N   & 0.74   &  -1.796 \\
6021.843 & Mn I                             & 3.08   &  -1.169 \\
6021.846 & Mn I                             & 3.08   &  -1.676 \\
6021.847 & Mn I                             & 3.08   &  -1.391 \\
6021.876 & $^{12}$C$^{14}$N   & 0.84   &  -2.202 \\
6021.880 & $^{12}$C$^{14}$N   & 1.72   &  -1.907 \\
6021.946 & Cr I                               & 3.85   &  -3.419 \\
6022.102 & Mo I                              & 3.26   &  -1.473 \\
\enddata

\end{deluxetable}

\clearpage

\begin{deluxetable}{ccccccc}
\setcounter{table}{2}
\tablewidth{400pt}
\tablecaption{Non-LTE corrections}
\tablehead{
\colhead{Star} &
\colhead{T$_{\rm eff}$(K)} &
\colhead{Log g} &
\colhead{[Fe/H]} &
\colhead{$\Delta$$_{\rm Non-LTE}$(6013\AA)} & 
\colhead{$\Delta$$_{\rm Non-LTE}$(6021\AA)}
}
\startdata
ROA213 &  4500 & 1.0  & -2.0 & +0.30 & +0.32 \\
ROA219 &  3900 & 0.7  & -1.4 & +0.12 & +0.06 \\
ROA236 &  4200 & 0.7  & -1.2 & +0.15 & +0.09 \\
ROA324 &  4000 & 0.7  & -1.0 & +0.11 & +0.05  \\
M4 2617 & 4280 & 1.2  & -1.0 & +0.11 & +0.04 \\
\enddata

\tablecomments{$\Delta$$_{\rm Non-LTE}$= A$_{\rm Non-LTE}$ - A$_{\rm LTE}$. 
}
\end{deluxetable}

\clearpage

\begin{figure}\plotone{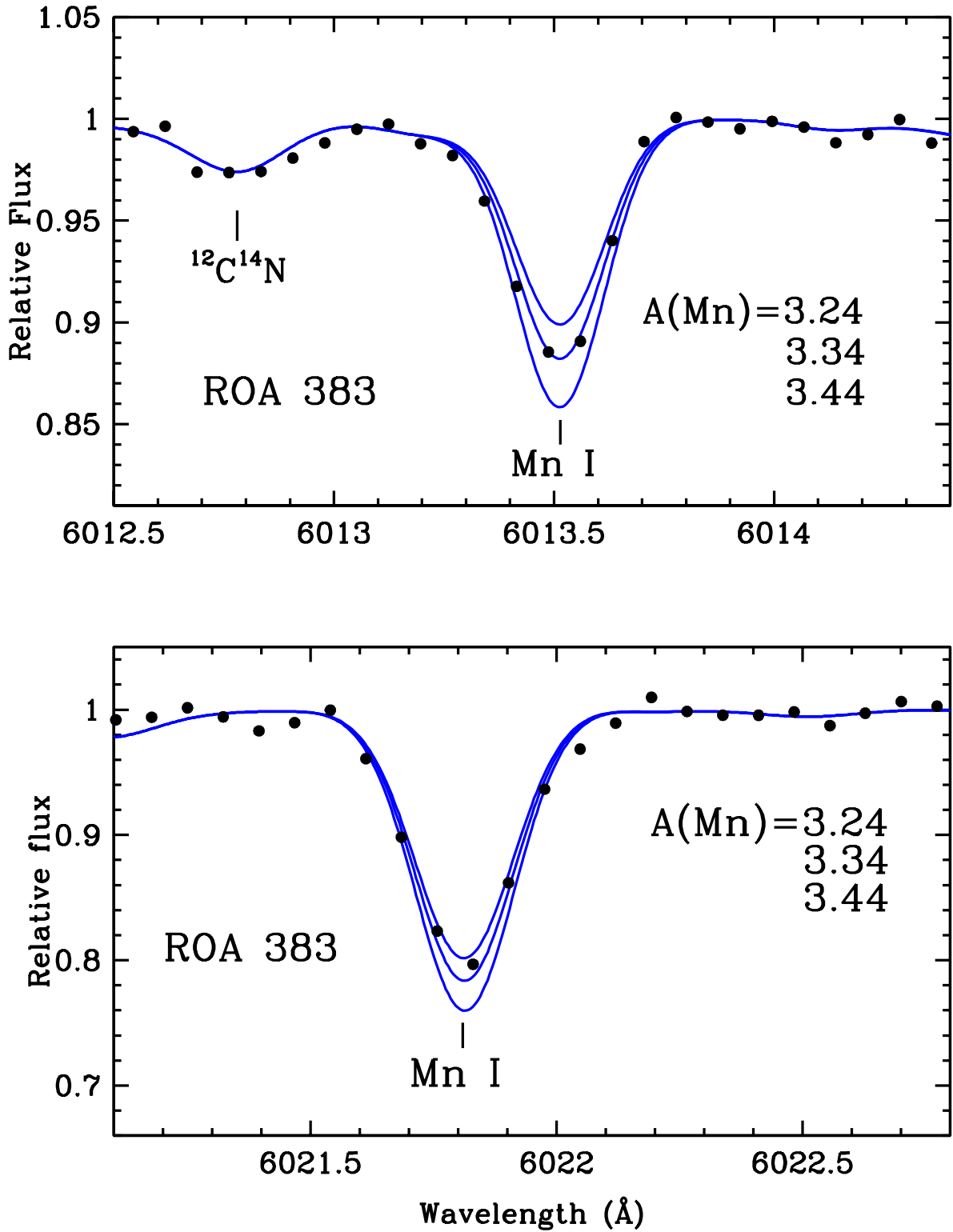}
\figcaption[f1.ps]{ Sample observed spectra (dotted lines) in the region of the Mn I lines in
target star ROA 383.
Synthetic spectra (solid lines) were computed for 3 different Mn abundances (A(Mn)= 3.24; 3.34; 3.44). 
\label{fig1}}\end{figure}
                                                                                                                      
\clearpage

\begin{figure}
\resizebox{\columnwidth}{!}{\includegraphics{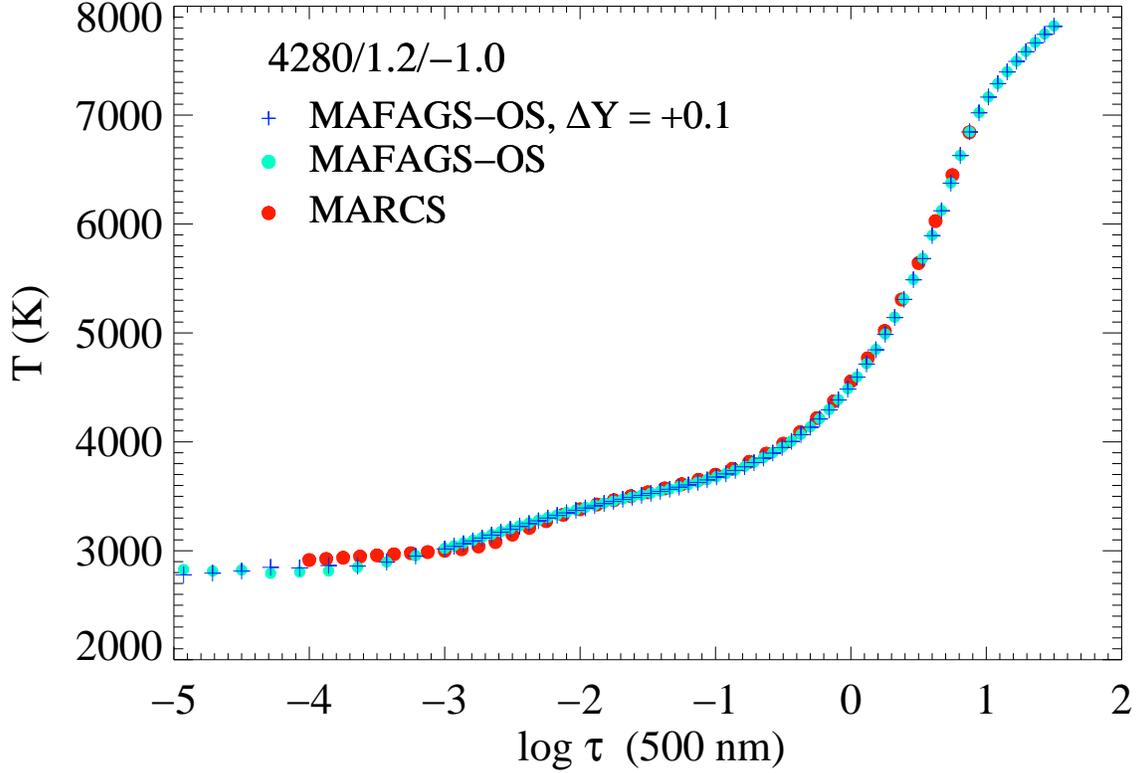}}
\vspace{0mm}
\caption{A comparison between the temperature stratification in the MAFAGS-OS model atmospheres 
(used to compute non-LTE corrections) with the model atmospheres adopted in the LTE analysis
(Bell et al. 1976). We also
show a model atmosphere computed with He mass fraction increased by 0.1 ($\Delta Y = +0.1$) 
relative to the scaled-solar value ($\Delta Y = 0$). 
The models correspond to T$_{eff}$ = 4280 K, log g = 1.2, [Fe/H] = -1.0.}
\label{temperature}
\end{figure}

\begin{figure}
\resizebox{9.0cm}{!}{\includegraphics{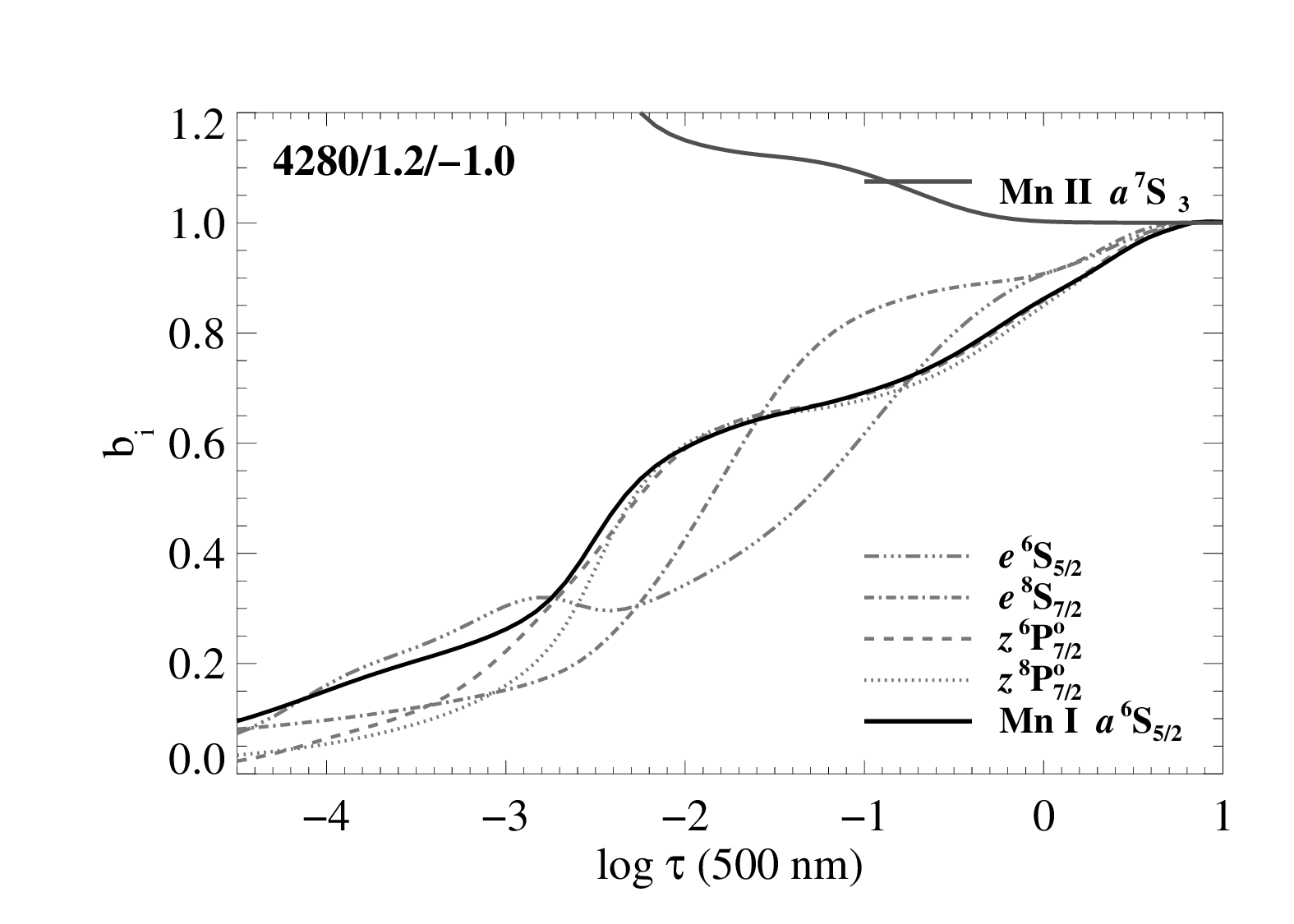}}
\resizebox{9.0cm}{!}{\includegraphics{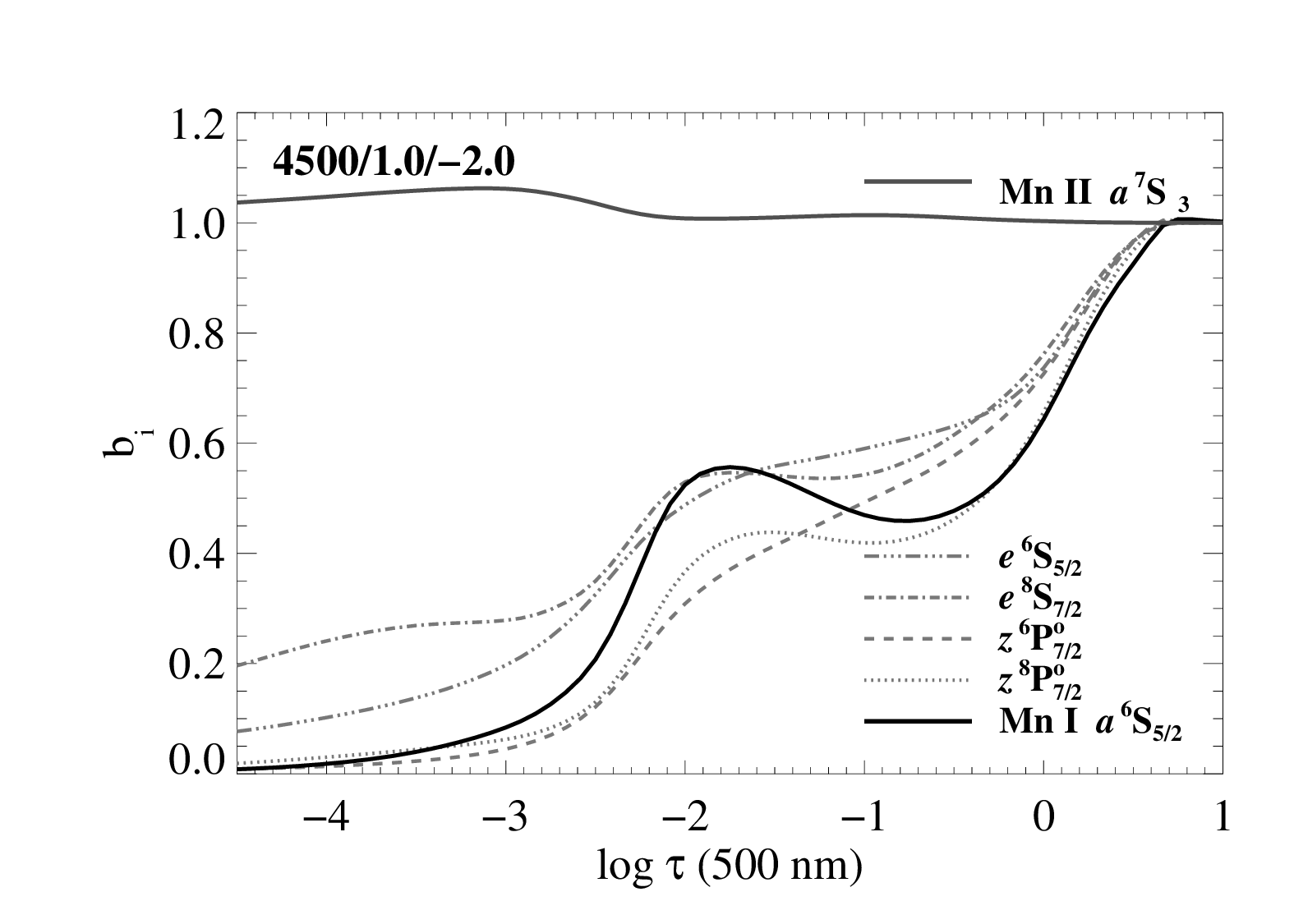}}
\vspace{0mm}
\caption{Departure coefficients $b_i$ of selected Mn I levels as a
function of continuum optical depth at $5000$ \AA. The model atmospheres are
computed with stellar parameters specified in each panel.}
\label{departures}
\end{figure}

\begin{figure}
\resizebox{13.0cm}{!}{\includegraphics{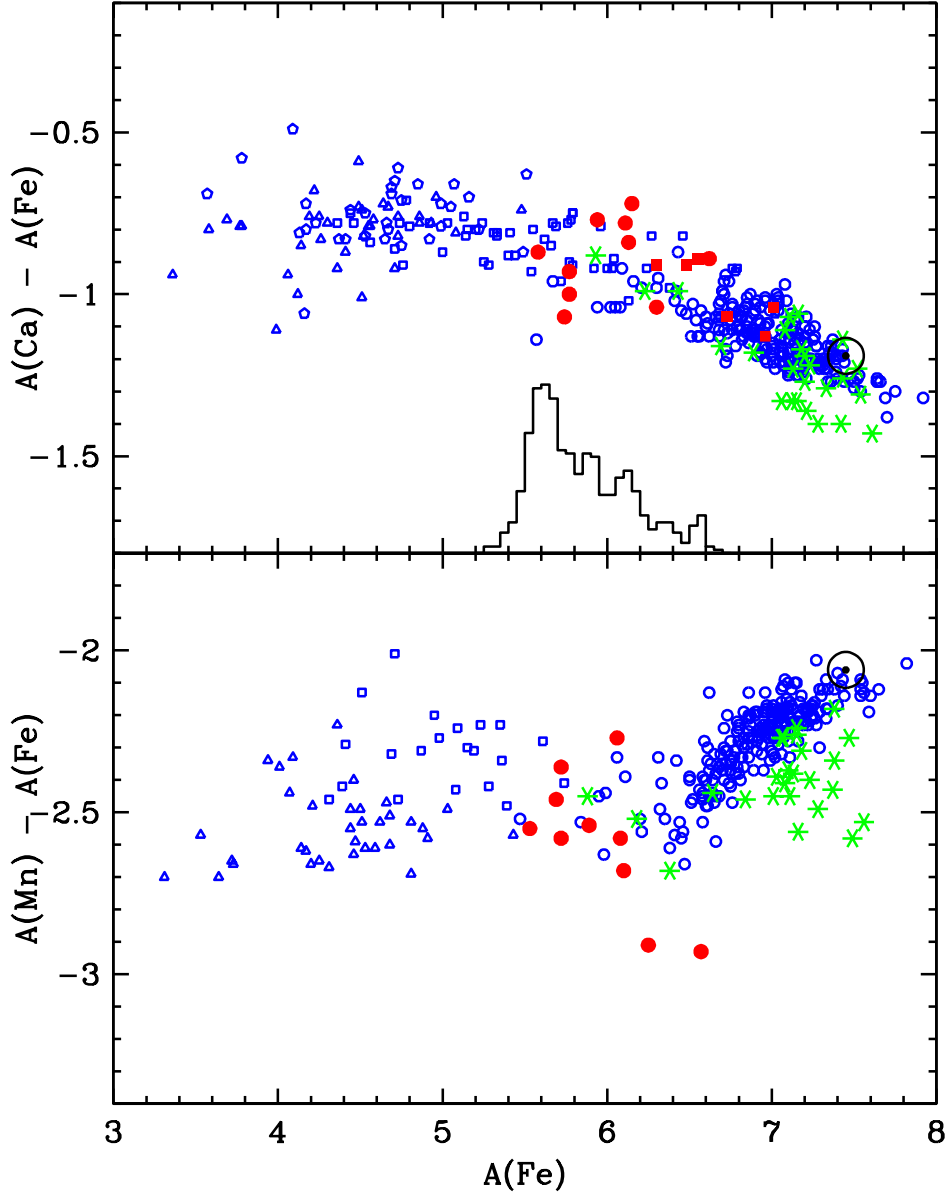}}
\figcaption[f2.ps]{LTE calcium (top panel) and  manganese (bottom panel) abundance results for $\omega$ Cen
in comparison with the Milky Way. The abundances for sample $\omega$ Cen stars 
(red filled circles) are from this study and Smith et al. (2000). 
Calcium abundances for 6 additional targets in $\omega$ Cen are from Pancino et al. (2002; red filled squares). 
The samples of stars representing the Milky Way disk and halo were taken from Reddy et al. (2003; 2006; blue open circles);
Fulbright (2002; blue open squares); Johnson (2002; blue open squares); Cayrel et al. (2004;
blue open triangles) and Mcwilliam et al. (1995; blue open pentagons). 
Abundance results for the Sagittarius dwarf galaxy by McWilliam et al. (2003) and Sbordone et al. (2007) are also shown 
(green asterisks). The metallicity distribution shown (top panel) is from Sollima et al. (2005).
\label{fig1}}\end{figure}

\clearpage

\begin{figure}
\resizebox{13.0cm}{!}{\includegraphics{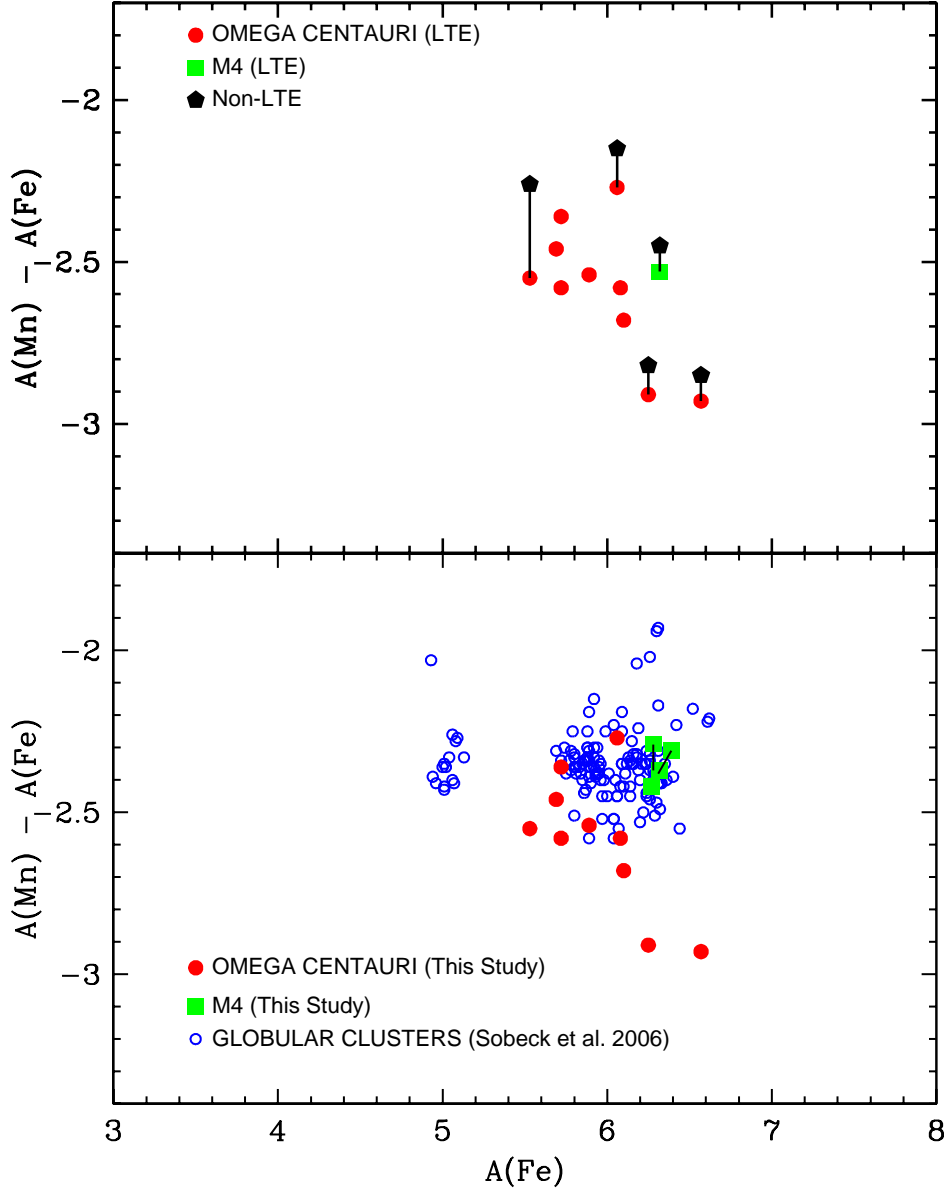}}
\figcaption[f3.ps]{Top panel: LTE Mn results for $\omega$ Cen (filled red circles) 
are shown along with the non-LTE corrections obtained for 5 targets. Note that the low [Mn/Fe] abundance of those stars
representative of the more metal rich population in $\omega$ Cen hold the same pattern in non-LTE. 
Bottom panel: The same LTE results for $\omega$ Cen from top panel (red filled circles)    
shown in comparison with abundances derived for all globular clusters studied in Sobeck et al. (2006; blue open circles).  
M4 results are also shown (as filled green squares) and these emphasize the differences between 
Mn abundances in M4 and $\omega$ Cen at a similar metallicity. 
Two M4 stars in this study are in common with the Sobeck et al. (2006) study and the derived Mn
abundances are in good agreement within the unceratinties (connected by black solid lines). The Mn abundances in 
Sobeck et al. (2006) have been adjusted by 0.16 dex in order to be on the same log gf scale as the results in this study.
\label{fig1}}\end{figure}                                                                                                                      
\clearpage

\end{document}